\documentclass[12pt,twocolumn,a4paper]{article}

\usepackage{times}
\usepackage{graphicx}

\title{Relativistic Positioning Systems: Numerical Simulations}
\author{Diego S\'aez\thanks{Corresponding author. E-mail: Diego.Saez@uv.es}, and Neus Puchades
\\ \\ \small{Departamento de Astronom\'{\i}a y Astrof\'{\i}sica, Universidad  de Valencia, 
46100 Burjassot, Valencia, Spain} }

\begin{document}
\label{paper:saez}

\maketitle

\begin{abstract}
The motion of satellite constellations similar to GPS and Galileo is 
numerically simulated and, then, the region where bifurcation (double positioning)
occurs is appropriately represented. In the cases of double positioning, 
the true location may be found using
additional information (angles or times). The zone where the Jacobian, J, of the
transformation from inertial to emission coordinates vanishes is also represented
and interpreted. It is shown that the uncertainties in the satellite world lines
produce positioning errors, which depend on the value of $|J|$. The smaller this quantity
the greater the expected positioning errors. Among all the available 4-tuples of satellites, 
the most appropriate one --for a given location--  should minimize positioning errors 
(large enough $|J|$ values) avoiding bifurcation. Our study is particularly important 
to locate objects which are far away from Earth, e.g., satellites.
\end{abstract}

\section{General considerations}
\label{sec1}

Global Navigation Satellite Systems (GNSSs) are satellite 
constellations broadcasting signals, which may be used to
find the position of a receiver (user)~\cite{coll12}.
GNSSs are mainly used to 
locate receivers on Earth surface. In this case, 
users receiving signals from four or more satellites 
may be always located with admissible positioning errors.
However, if the user to be located is far away from Earth, two main problems
may arise, the first one is the existence of two possible positions for the object 
(bifurcation), some previous considerations about this problem may be found in~\cite{abe91,cha94,gra96}.
The second problem is related to the positioning errors 
due to uncertainties in the world lines of the satellites. These 
errors are proved to be too big for some user positions close to points 
where the Jacobian $J$ --of the 
transformation from inertial to emission coordinates-- is too small.
In this paper, we find and represent the regions where the 
above problems arise. Only numerical calculations may be used
to find these regions in the case 
of realistic GNSSs. Moreover, their representation is 
an additional problem, which have been solved by choosing 
appropriate sections --of the 4D emission region-- 
and by using suitable methods previously developed in other 
research fields (see below).
The mentioned regions have been only studied in the interior
of a big sphere having a radius of $10^{5} \ km $, which is 
centred in a point E of the Earth surface (see Fig.~\ref{fig1}).

Quantities $G$, $M_{\oplus}$, $t$, and $\tau $ stand for the gravitation constant,
the Earth mass, the coordinate time, and the proper time, respectively.
Greek (Latin) indexes run from $0$ to $3$ ($1$ to $3$). Quantities 
$\eta_{\alpha \beta}$ are the covariant components of the Minkowski 
metric tensor. Our signature is (+,+,+,--). The unit of distance is assumed to be 
the Kilometre, and the time unit is chosen 
in such a way that the speed of light is $c=1$. The index $A$ numerates the
four satellites necessary for relativistic positioning.

GPS and GALILEO satellite constellations are simulated~\cite{puc11,puc12}.
Satellite trajectories are assumed to be circumferences 
in the Schwarzschild space-time
created by an ideal spherically symmetric Earth.
A first order approximation in $GM_{\oplus}/R$ 
is sufficient for our purposes.
The angular velocity is $\Omega = (GM_{\oplus}/R^{3})^{1/2} $,  
and coordinate and proper times are related as follows:
$\gamma = \frac {dt}{d\tau} = ( 1 - \frac {3GM_{\oplus}}{R})^{-1/2} \ . $
Angles $\theta$ and $\phi$ fix the orbital plane
(see~\cite{puc12}),
and the angle $\alpha_{A}(\tau) = \alpha_{A0} - \Omega \gamma \tau $ 
localizes the satellite on its trajectory. 
This simple model is good enough as
a background configuration. Deviations with respect to
the background satellite world lines will be necessary to develop
our study about positioning accuracy (see below).

Other known world lines (no circumferences) 
of Schwarzschild space-time might be easily implemented in the code, but the
new background satellite configurations would lead to 
qualitatively comparable numerical results; at least,  
for the problems considered in this paper.

The angle $\alpha_{A}(\tau)$ may be calculated for every
$\tau $, and the two angles fixing the orbital plane 
($\theta $ and $\phi $) are constant.
From these three angles and the proper times $\tau^{A} $, the inertial coordinates
of the four satellites, $x^{\alpha}_{A}$,
may be easily found to first order in $GM_{\oplus}/R$~\cite{puc12}.
This means that the world lines of the background satellites 
[functions $y^{\alpha} = x^{\alpha}_{A}(\tau^{A})$] are known for 
every satellite $A$. Hence,
given the emission coordinates 
$( \tau^{1},\tau^{2},\tau^{3},\tau^{4})$ of a receiver, the inertial coordinates
$x^{\alpha}_{A} \equiv (x_{A},y_{A},z_{A},t_{A})$
of the four satellites --at emission times-- may be easily calculated. 
The knowledge of the satellite world lines is
necessary for positioning; namely, to find the inertial coordinates
from the emission ones~\cite{coll10}. 

The satellite world lines are also 
necessary to go from the inertial coordinates of an user to its emission ones.
This transformation is now considered under the assumption that
photons move in the Minkowski space-time, whose 
metric has the covariant components $\eta_{\alpha \beta}$. This approach
is good enough for us.
Since photons follow null geodesics from emission 
to reception, the following algebraic equations must be satisfied:
\begin{equation}
\eta_{\alpha \beta} [x^{\alpha} - x^{\alpha}_{A}(\tau^{A}) ]  
[ x^{\beta} - x^{\beta}_{A}(\tau^{A}) ] = 0 \ . 
\label{dn1}
\end{equation}
These four equations must be numerically solved to get the four emission 
coordinates $\tau^{A} $.
The four proper times are the unknowns in the system  (\ref{dn1}), 
which may be easily solved by using the well known Newton-Raphson method~\cite{pre99}.
Since the satellite world lines are known, functions $x^{\alpha}_{A}(\tau^{A}) $
may be calculated for any set of proper times $\tau^{A}$, thus, the
left hand side of Eqs.~(\ref{dn1}) can be computed and, consequently, 
the Newton-Raphson method may be applied.
A code has been designed to implement this method. It requires 
multiple precision. Appropriate tests have been performed~\cite{puc12}.

Moreover, given four emission coordinates $\tau^{A} $, Eqs.~(\ref{dn1}) could  be 
numerically solved to get the unknowns $x^{\alpha} $, that is to say, the 
inertial coordinates (positioning); however, this numerical method is not used. 
It is better the use of a certain analytical
formula giving $x^{\alpha} $ in terms of $\tau^{A}$, which was derived in~\cite{coll10}. 
The analytical formula is preferable because of the following reasons:
(i) the numerical method based on Eqs.~(\ref{dn1}) is more time consuming and,
(ii) the analytical formulation of the problem allows us a systematic 
and clear discussion of bifurcation, and also a study of the 
positioning errors close to points of vanishing Jacobian.

The analytical formula~\cite{coll10} has been 
described in various papers~\cite{coll10,coll12,coll11,puc12},
and numerically applied in~\cite{puc11,puc12}.  
This formula involves function $\chi^{2}$, which is the modulus of the 
configuration vector, and a discriminant $\Delta $. The definitions of 
both $\chi^{2}$ and $\Delta $ may be found in~\cite{coll10,coll12, coll13}. 
It is very important that these two quantities may be calculated 
by using only the emission coordinates 
$\tau^{A}$.
From the analytical formula giving the inertial coordinates 
in terms of the emission ones,
and taking into account some basic relations of Minkowski space-time,
the following propositions have been previously proved~\cite{coll10,coll11,coll12}:

(a) For $\chi^{2} \leq 0$, there is only a positioning (past-like) solution.

(b) For $\chi^{2} > 0$ there are two positioning solutions; namely, there 
are two sets of inertial coordinates (two physical real receivers) associated 
to the same emission coordinates $\tau^{A}$.

(c) The Jacobian $J$ of the transformation giving the emission coordinates 
in terms of the inertial ones vanishes if and only if the discriminant $\Delta $ vanishes.

(d) The Jacobian $J$ may only vanish if $\chi^{2} > 0 $; namely, in the 
region of double positioning (bifurcation).

(e) The Jacobian $J$ may only vanish if the lines of sight --at emission times-- 
of the four satellites belong to the same cone (with vertex in the user).

These conclusions are basic for the numerical estimations and discussions 
presented below. In particular, after calculating $\chi^{2}$ and $\Delta$ from the 
emission coordinates, propositions (a) and (b)
allow us to find the regions where bifurcation takes place, 
whereas the zones with vanishing Jacobian (infinite positioning errors) may be found by using 
proposition (c).

\section{Emission region}
\label{sec2}

\begin{figure*}[ht]
\centering
\includegraphics[width=11cm,angle=90]{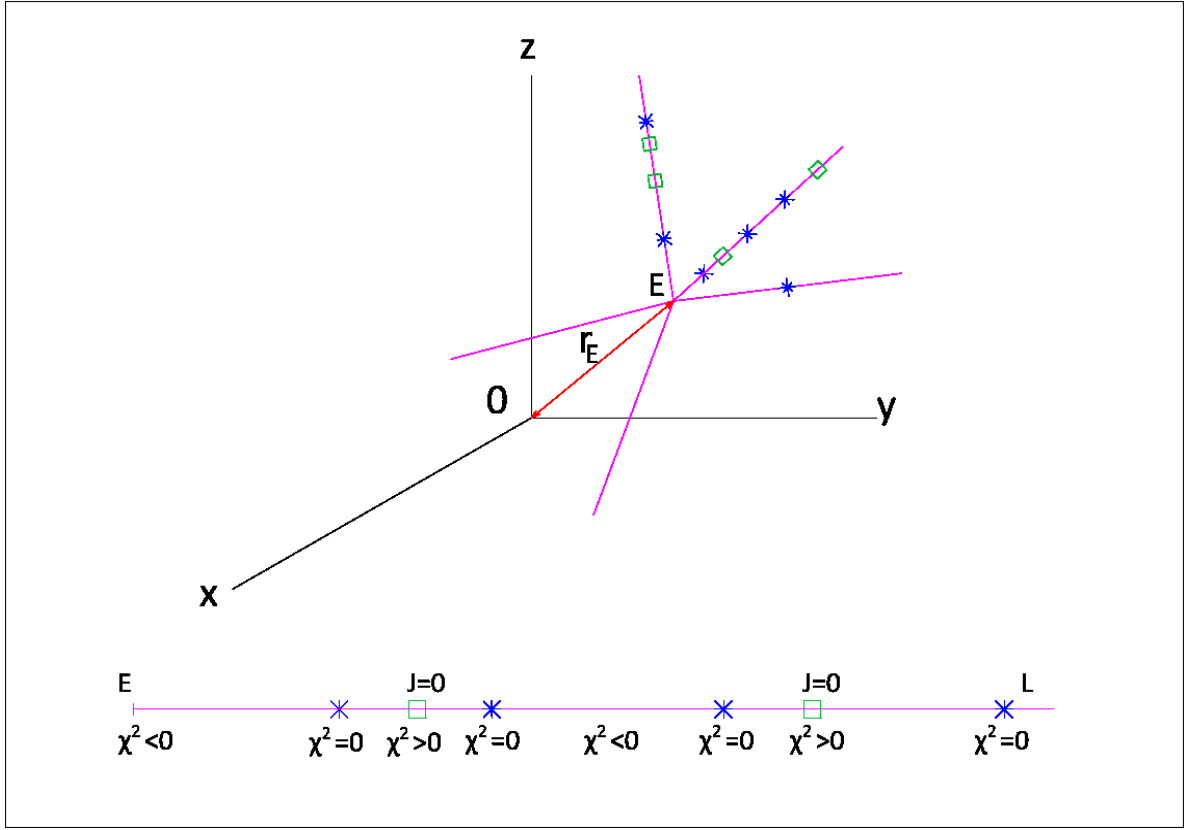}
\caption{3D sections ($t=constant$) of the  emission region  
are considered. Point E is an arbitrary centre. Its
distance to the origin $O$ is the Earth radius. $L$ is the distance from E 
to another point of the section.
3072 segments starting from E and having distinct directions cover each 3D section. 
Along each direction
our study is restricted to $0<L<L_{max} = 10^{5} \ km$. 
We look for the zeros
of $\chi^{2} $ and $J$. The Jacobian only may vanish in the segments 
where $\chi^{2} > 0$, which are limited by the first and second or by  
the third and fourth $\chi^{2} $-zeros.}\label{fig1}
\end{figure*}

For every set of four satellites, 
the so-called 4D emission region~\cite{coll10,coll12,puc12} 
is studied by considering 3D sections $t=constant$. 
Each of these sections is covered by points according to the 
method described in Fig.~\ref{fig1}. Let us here give some 
additional details. Healpix package --first used in cosmic microwave background researches~\cite{gor99}-- 
is used to
define 3072 directions. A segment 
with one of its ends at $E$ and having length $L_{max} = 10^{5} \ km$ 
is associated to each direction. 
A great enough number of points are uniformly distributed  
along every direction. The inertial coordinates of the 
chosen points are known by construction; then, the 
Newton-Raphson method --implemented in our numerical codes--
gives the associated emission coordinates, which 
allow us the computation of $\chi^{2} $ and $\Delta $ (see section~\ref{sec1}). 
With these quantities, we may find the zeros of both 
$\chi^{2} $ and $J$ [proposition (c) of section~\ref{sec1}] 
along any given direction.
As it is displayed in Fig.~\ref{fig1}, 
various zeros of $\chi^{2} $ and $J$ may be found 
in each direction. They may be distributed in different 
ways by obeying proposition (d) of section \ref{sec1}.

Once the zeros of $\chi^{2} $ and $\Delta $ have been found 
for all the Healpix directions (3072), an appropriate 
method is necessary to display the results. 
Our method is based 
on the Healpix pixelisation and the mollwide projection (see~\cite{puc12}
for more details). Healpix associates a pixel to each direction. 
The pixel colour measures the value of some chosen quantity according to
the colour bar displayed in our figures, which are mollwide projections of the 
pixelised sphere. 

\begin{figure*}[t] 
\centering 
(a) 
{\label{fig:a}\includegraphics[width=0.45\textwidth,height=0.42\columnwidth] 
{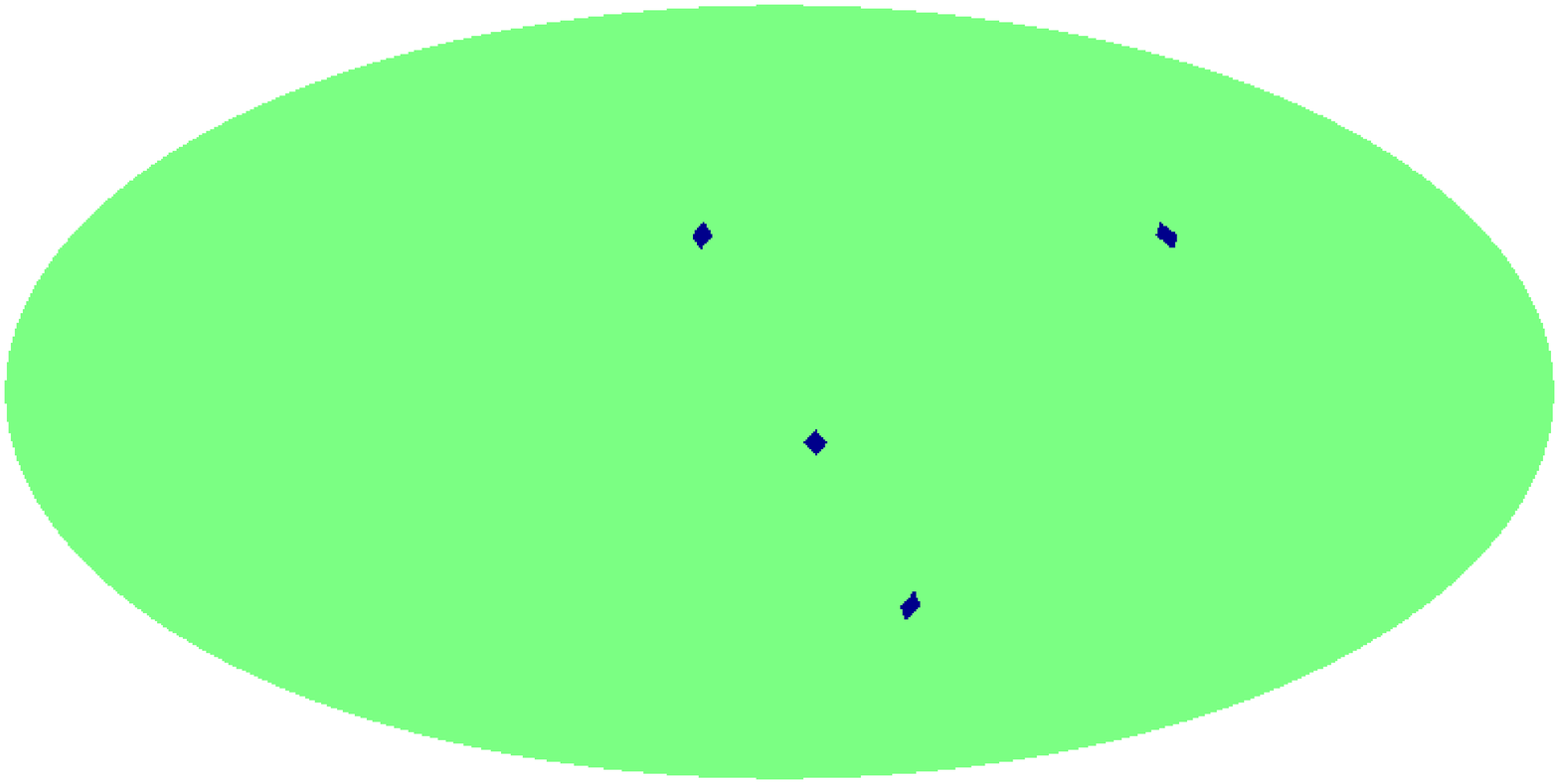}} 
\\
(b) 
{\label{fig:b}\includegraphics[width=0.45\textwidth,height=0.42\columnwidth] 
{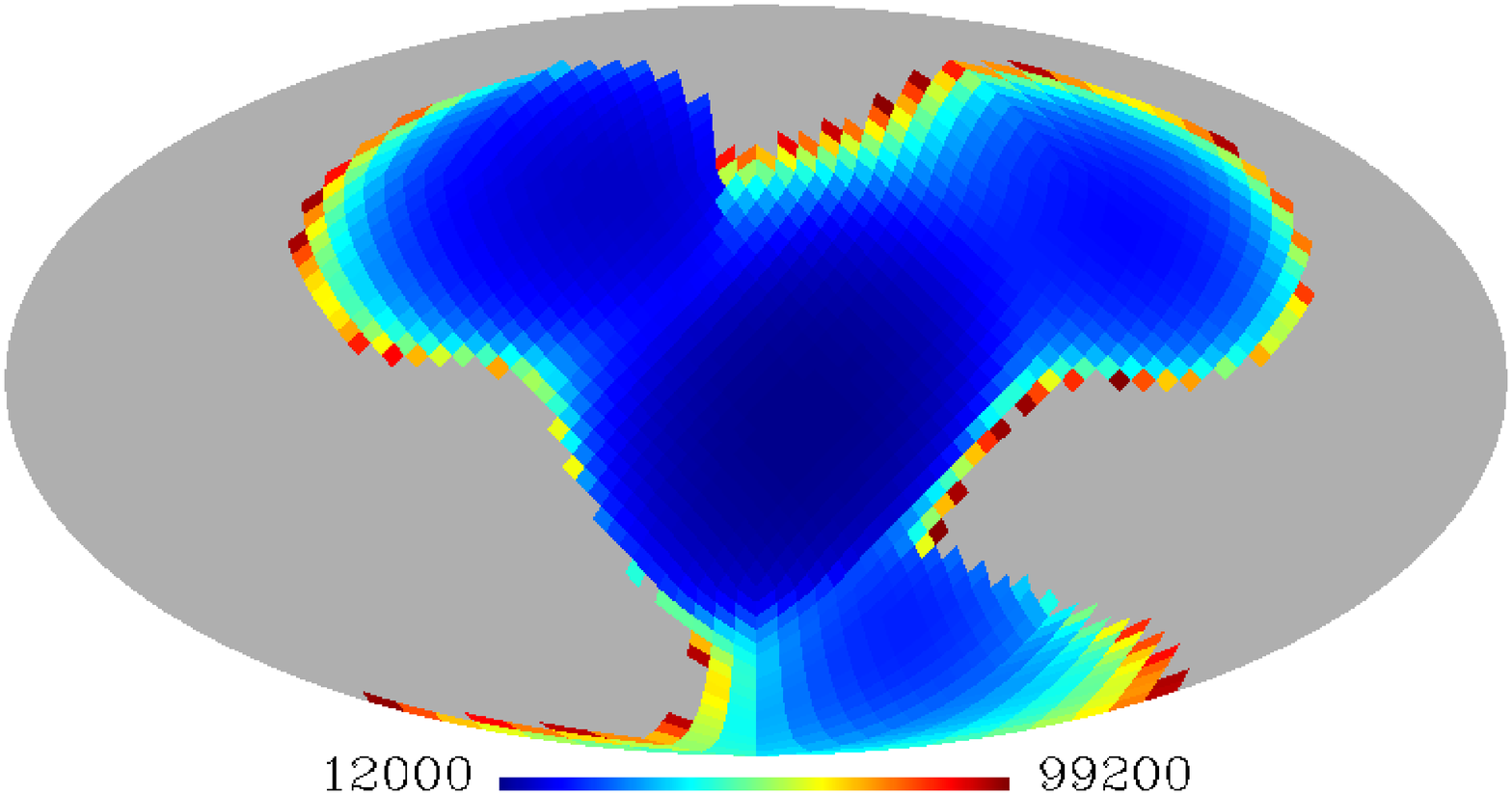}} 
(e) 
{\label{fig:e}\includegraphics[width=0.45\textwidth,height=0.42\columnwidth] 
{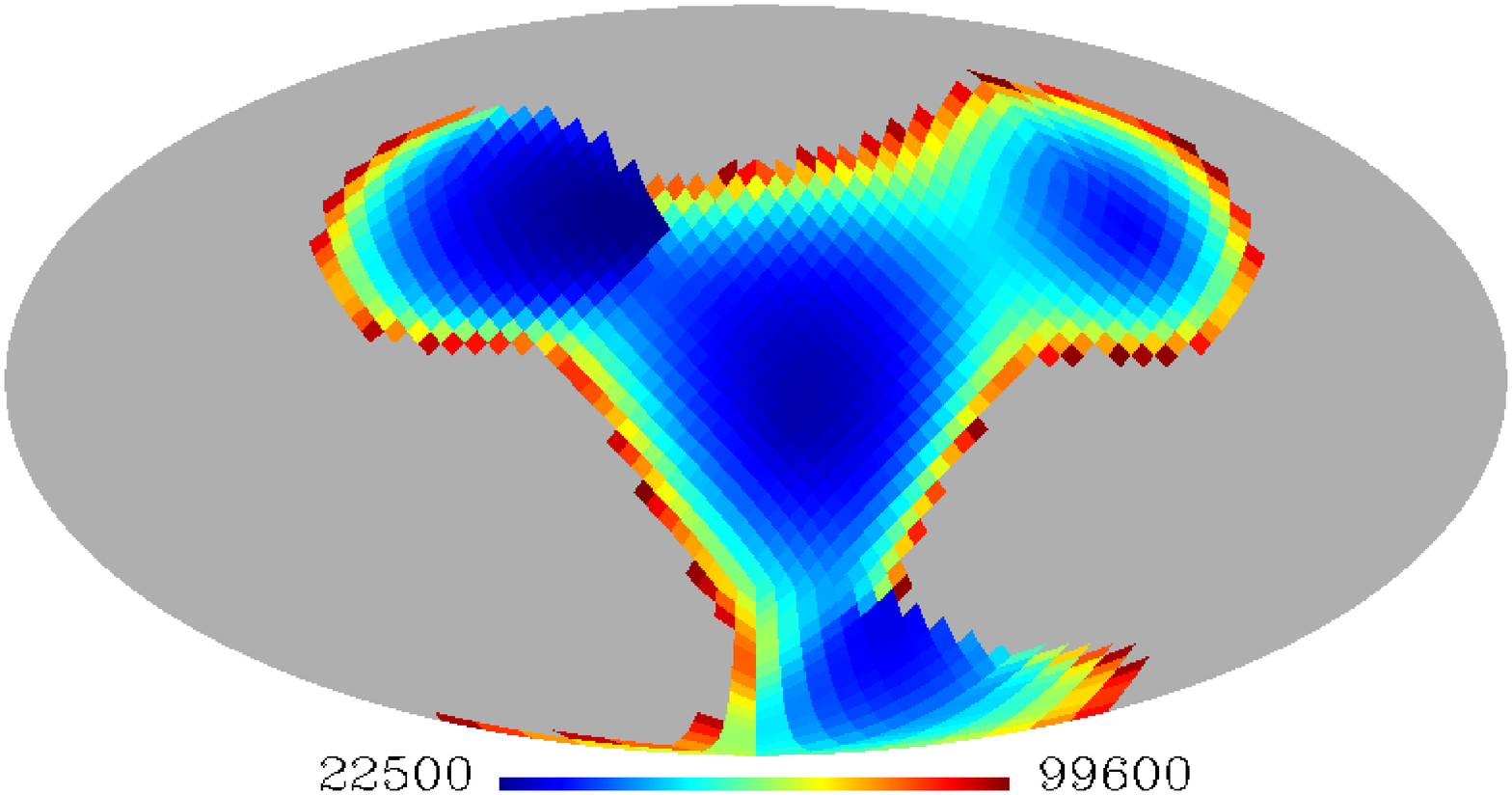}}  
\\
(c) 
{\label{fig:c}\includegraphics[width=0.45\textwidth,height=0.42\columnwidth] 
{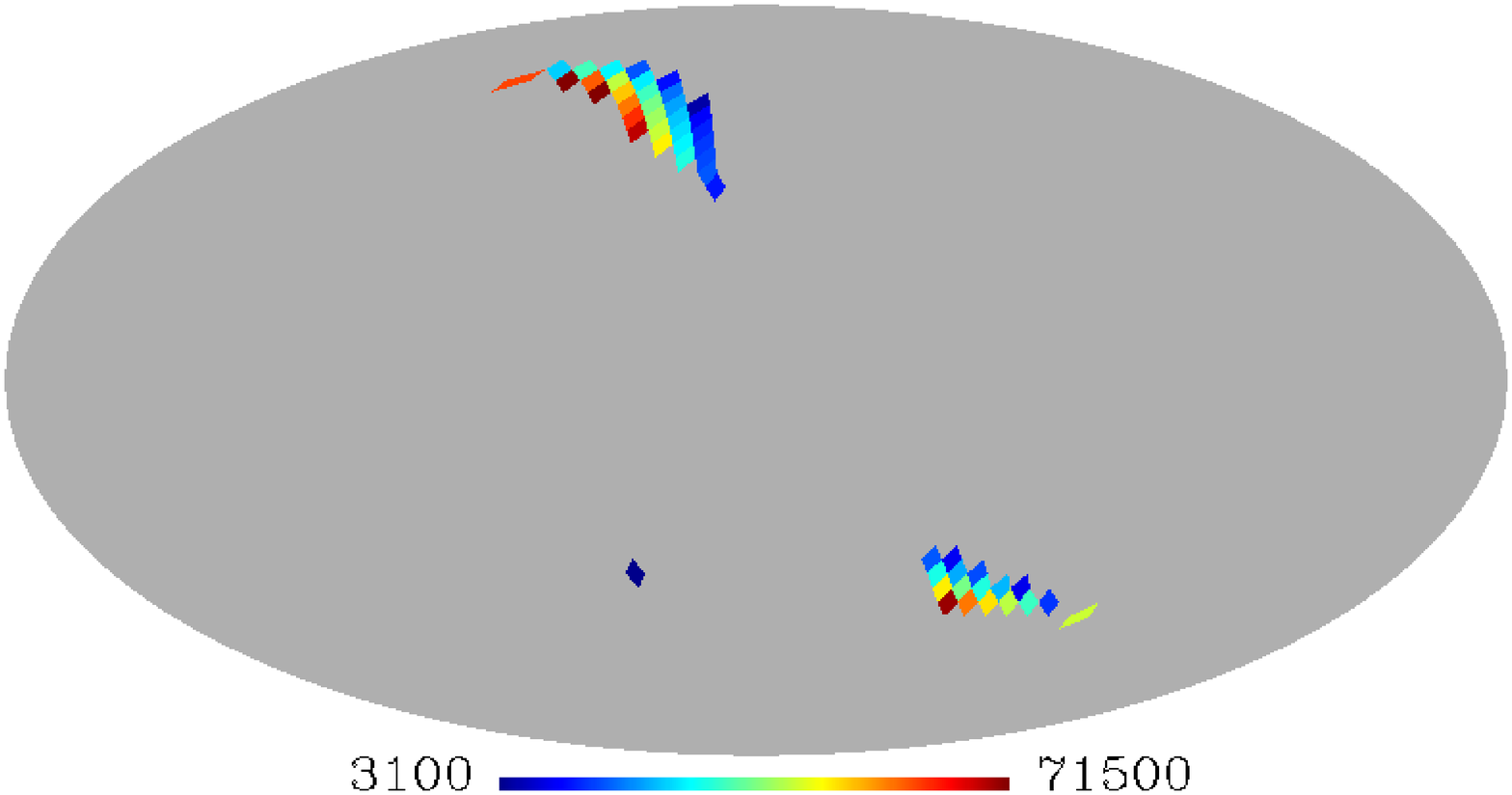}} 
(f) 
{\label{fig:f}\includegraphics[width=0.45\textwidth,height=0.42\columnwidth] 
{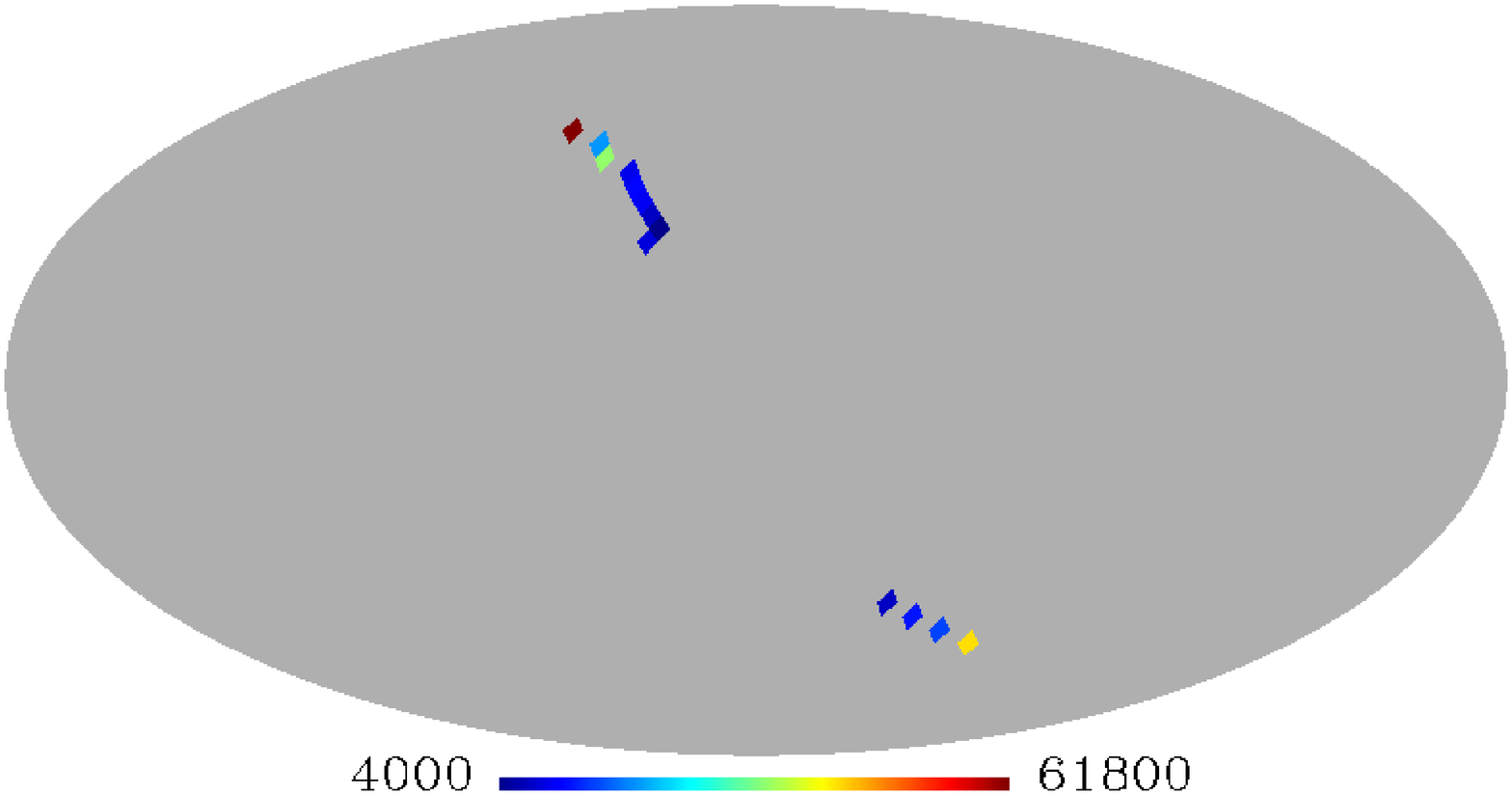}} 
\\
(d) 
{\label{fig:d}\includegraphics[width=0.45\textwidth,height=0.42\columnwidth] 
{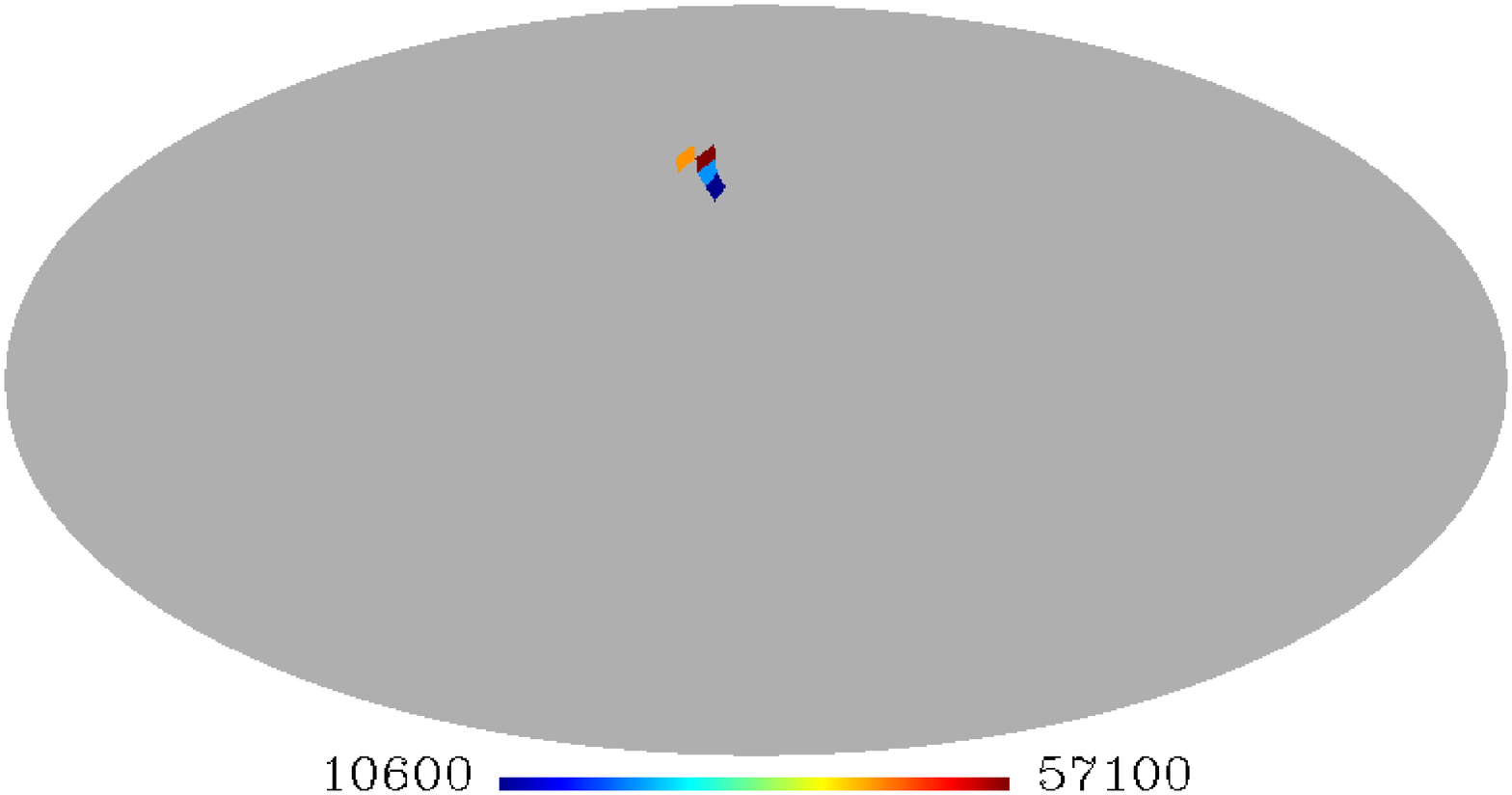}} 
(g) 
{\label{fig:g}\includegraphics[width=0.45\textwidth,height=0.42\columnwidth] 
{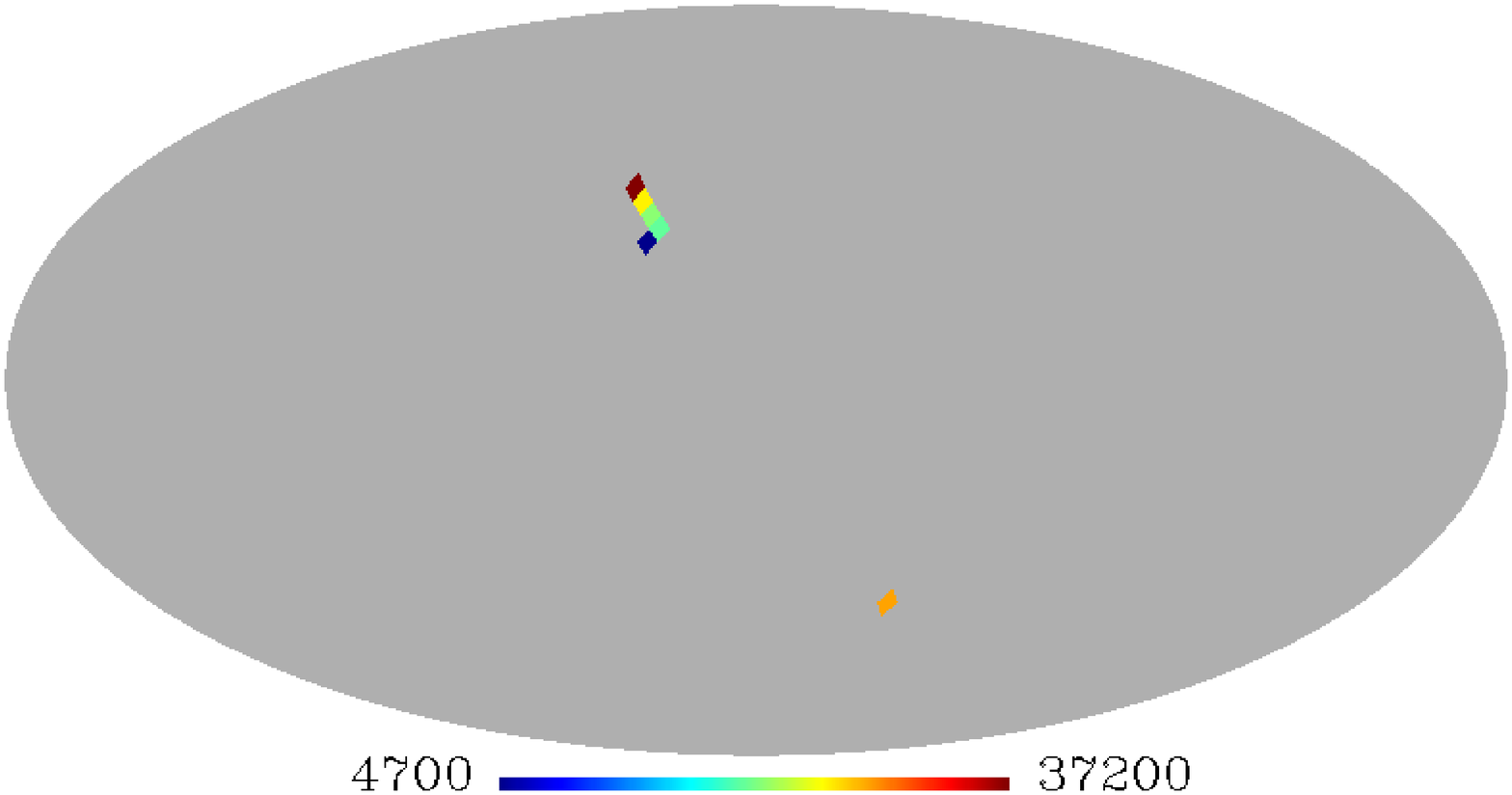}} 
\caption{Structure of section $x^{4} = t = 25 \ h $ for $L < 10^{5} \ km $ 
} 
\label{fig2} 
\end{figure*}

The structure of the 3D section considered in Fig.~\ref{fig2}
($t= 25 \ h$ after the time origin) is displayed in seven panels.
The blue pixels of panel~(a) show the directions of the 
four chosen satellites when they emitted the signals received 
at point E. Since satellite velocities are much smaller 
than the speed of light, the satellite positions at emission 
times are very similar for every point
of the 3D section under consideration.
In order to understand
panels (b)--(h) the reader need to know the quantity associated to  
every colour bar and the meaning of the grey pixels.
In panel (b) [(e)], the colour bar shows
the distance from $E$ 
to the first zero of $\chi^{2}$ [$J$], and along the 
directions corresponding to the grey pixels, 
function $\chi^{2}$ [$J$] does not vanish, at least, 
up to a distance $L_{max} $ from $E$.
In panel (c) [(f)], the colour bar gives the 
distance from the first to the second zero of $\chi^{2}$ [$J$],
and function $\chi^{2}$ [$J$] does not vanish two times
in the directions of the grey pixels (up to $L_{max} $) and, finally, 
in panel (d) [(g)] the colour bar displays distances 
from the second to the third zero 
of $\chi^{2}$ [$J$], and no a third zero of 
$\chi^{2}$ [$J$] has been found along the 
directions of the grey pixels (up to $L_{max} $).
In some cases, there are no directions with more than 
one zero of functions $\chi^{2}$ (see~\cite{puc12}) and $J$.
In any case, panels (c)--(g) of Fig.~\ref{fig2} show 
that there are only few directions with 
two zeros of these functions, and also that 
directions with three zeros are very scarce. Other 3D sections
and other 4-tuples of satellites have been studied with 
similar results.

From Figs.~(\ref{fig1}) and~(\ref{fig2}),
it follows that the emission 
region has the following structure: there are directions without
zeros of $\chi^{2}$ and $J$. These directions subtend a great
solid angle [grey pixels of panel.~(b) and~(e)].
The complementary solid angle corresponds to directions
with one or more zeros of $\chi^{2}$. From point E 
to the first zero of $\chi^{2}$ there is no bifurcation,
but it appears beyond the first zero. For the small number of
directions having a second zero, there is no
bifurcation beyond this zero, but it occurs again beyond the third zero (see Fig.~\ref{fig1}),
which only exists for
very scarce directions. We have verified that, according to 
proposition (d) of section~\ref{sec1}, the Jacobian only vanishes 
in regions with bifurcation.

\section{Positioning errors and satellite uncertainties}
\label{sec3}

\begin{figure*}[ht]
\centering
\includegraphics[width=7cm]{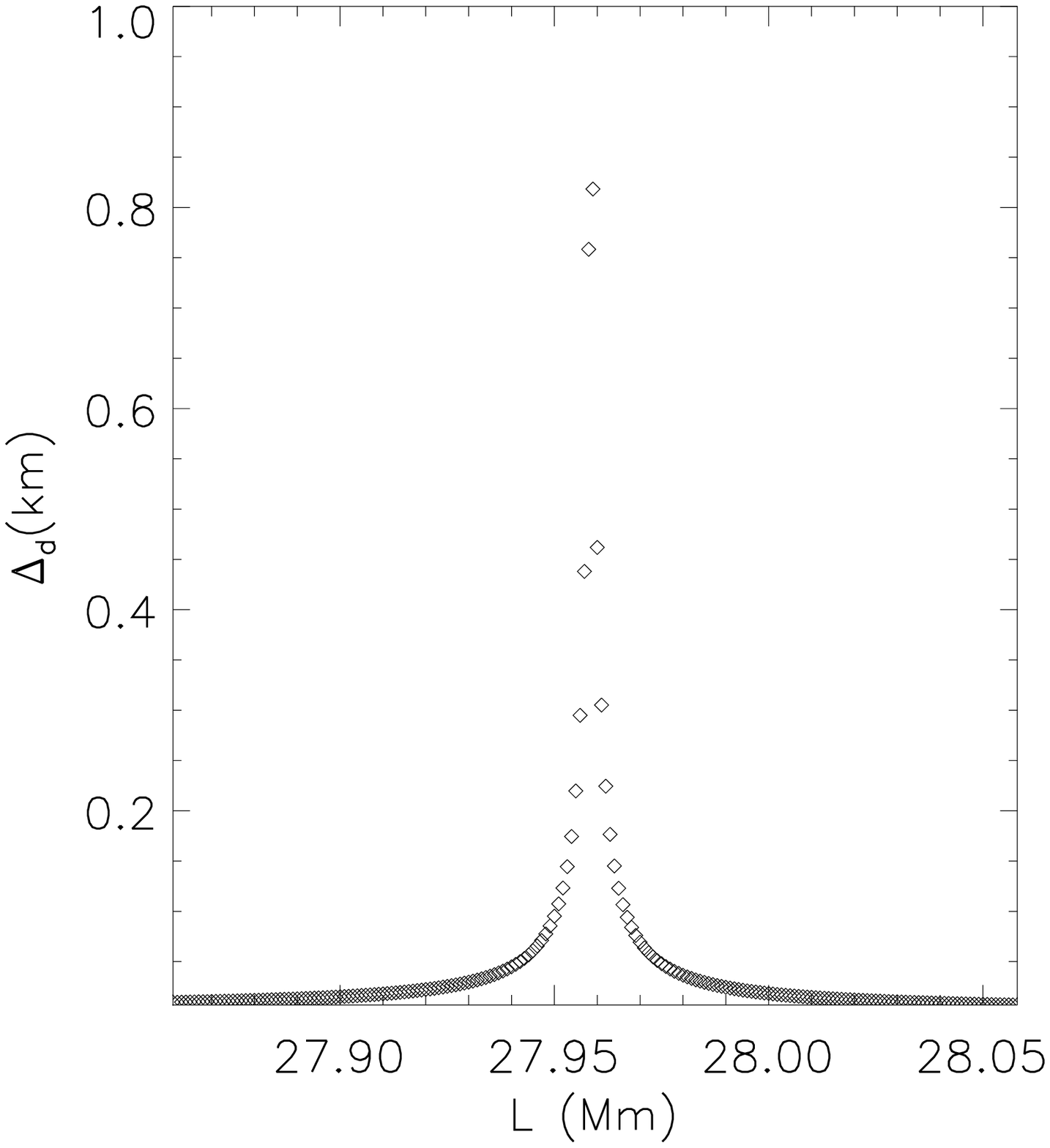}
\caption{Estimator $\Delta_{d}$ (in $km $) against the distance $L$ to $E$ (in $Mm\equiv10^{3}km$)
along the chosen direction in a $200 \ km$ interval centred at 
a point where $J=0 $. Close to this point, quantity $\Delta_{d}$
takes on very large values}\label{fig3}
\end{figure*}

The background world lines of the satellites are the circumferences 
of section \ref{sec1}, whose equations have the form
$y^{\alpha} = x^{\alpha}_{A}(\tau^{A})$. Let us first suppose  
that the background world lines are exactly followed by the 
satellites (without uncertainties). Under this assumption,        
given the inertial coordinates $x^{\alpha} $ of an user, 
the background world line equations,
Eqs.~(\ref{dn1}), the Newton-Raphson method, and multiple precision 
may be used to find the emission coordinates $\tau^{1},\tau^{2},\tau^{3},\tau^{4}$
with very high accuracy. Finally, the chosen inertial coordinates $x^{\alpha} $ may be recovered 
from the emission ones --with very high accuracy-- by using the analytical solution
derived in~\cite{coll10}. This process is useful to prove that 
our numerical codes work with high accuracy.

Let us now suppose that there are uncertainties in the satellite world 
lines, whose equations are $y^{\alpha} = x^{\alpha}_{A}(\tau^{A})+ \xi^{\alpha}_{A}$,
where $\xi^{\alpha}_{A}$ are deviations with respect to the background world lines
due to known or unknown external actions on the satellites. Let us now take 
the above inertial coordinates $x^{\alpha} $, the equations
$\eta_{\alpha \beta} [x^{\alpha} - x^{\alpha}_{A}(\tau^{A})- \xi^{\alpha}_{A}]  
[ x^{\beta} - x^{\beta}_{A}(\tau^{A}) -\xi^{\beta}_{A} ] = 0$, the Newton-Raphson method, and multiple precision,
to get the perturbed emission coordinates 
[$\tau^{1}+\Delta(\tau^{1}),\tau^{2}+\Delta(\tau^{2}),\tau^{3}+\Delta(\tau^{3}),\tau^{4}+\Delta(\tau^{4})$].
Since the time deviations $\Delta(\tau^{A})$ are all small, quantities $\xi^{\alpha}_{A}$
may be assumed to be constant in the short interval [$\tau^{A},\tau^{A}+ \Delta(\tau^{A})$].
Finally, 
by using the analytical solution mentioned above,
new inertial coordinates $x^{\alpha} + \Delta(x^{\alpha})$ may be obtained 
from the emission coordinates $\tau^{A}$ and the deviations $\xi^{\alpha}_{A}$.  
Coordinates  $x^{\alpha} + \Delta(x^{\alpha})$
are to be compared with the inertial coordinates $x^{\alpha} $ initially assumed.

Quantity $\Delta_{d} = [\Delta^{2}(x^{1})+\Delta^{2}(x^{2})+\Delta^{2}(x^{3})]^{1/2} $
is a good estimator of the positioning errors produced by the assumed uncertainties, 
$\xi^{\alpha}_{A}$, in the satellite motions.
         
For a certain direction, we have taken an interval of $200 \ km $ centred 
at a zero of function $J$ and, 
then, quantity $\Delta_{d} $ has been calculated in 200 uniformly 
distributed points of the chosen interval. 
In each of these points, the same deviations 
$\xi^{\alpha}_{A}$ have been used to perturb the satellite world lines.
The three quantities $\xi^{i}_{A}$ have been written in terms of 
the modulus $\Xi_{A} = [(\xi^{1}_{A})^{2} + (\xi^{2}_{A})^{2} + (\xi^{3}_{A})^{2}]^{1/2} $
and two angles $\Theta $ and $\Phi $ (spherical coordinates) and, then, 
quantities $\Xi_{A} $, $\Theta $, $\Phi $, and $\xi^{4}_{A}$ have been generated 
as random uniformly distributed numbers in the intervals [$0,10^{-3}$] in $km$, [$0,\pi$], [$0,2\pi$], and  
[$0,10^{-3}$] in time units (see section~\ref{sec1}), respectively. 
Results are presented in Fig.~\ref{fig3}, where we see that our
estimator of the positioning errors $\Delta_{d} $ is
very large close to the central point where $J=0$. 
This fact is important in section~\ref{sec4} (satellite positioning).
It is due to the fact that the satellite to be located 
may cross the region of vanishing $J$.

The Jacobian is being numerically calculated --in any point of the emission region--
to perform a relativistic study 
of the so-called {\em dilution of precision}~\cite{lan99}; namely,
to look for the relation between the geometry of the system 
satellites-user and the amplitude of the positioning errors. This study  
must be developed for users on Earth, as well as for users far away from Earth (satellites).

\section{Looking for the position of a satellite}
\label{sec4}

\begin{figure*}[ht] 
\centering 
(h) 
{\label{fig:h}\includegraphics[width=0.65\textwidth,height=0.7\columnwidth] 
{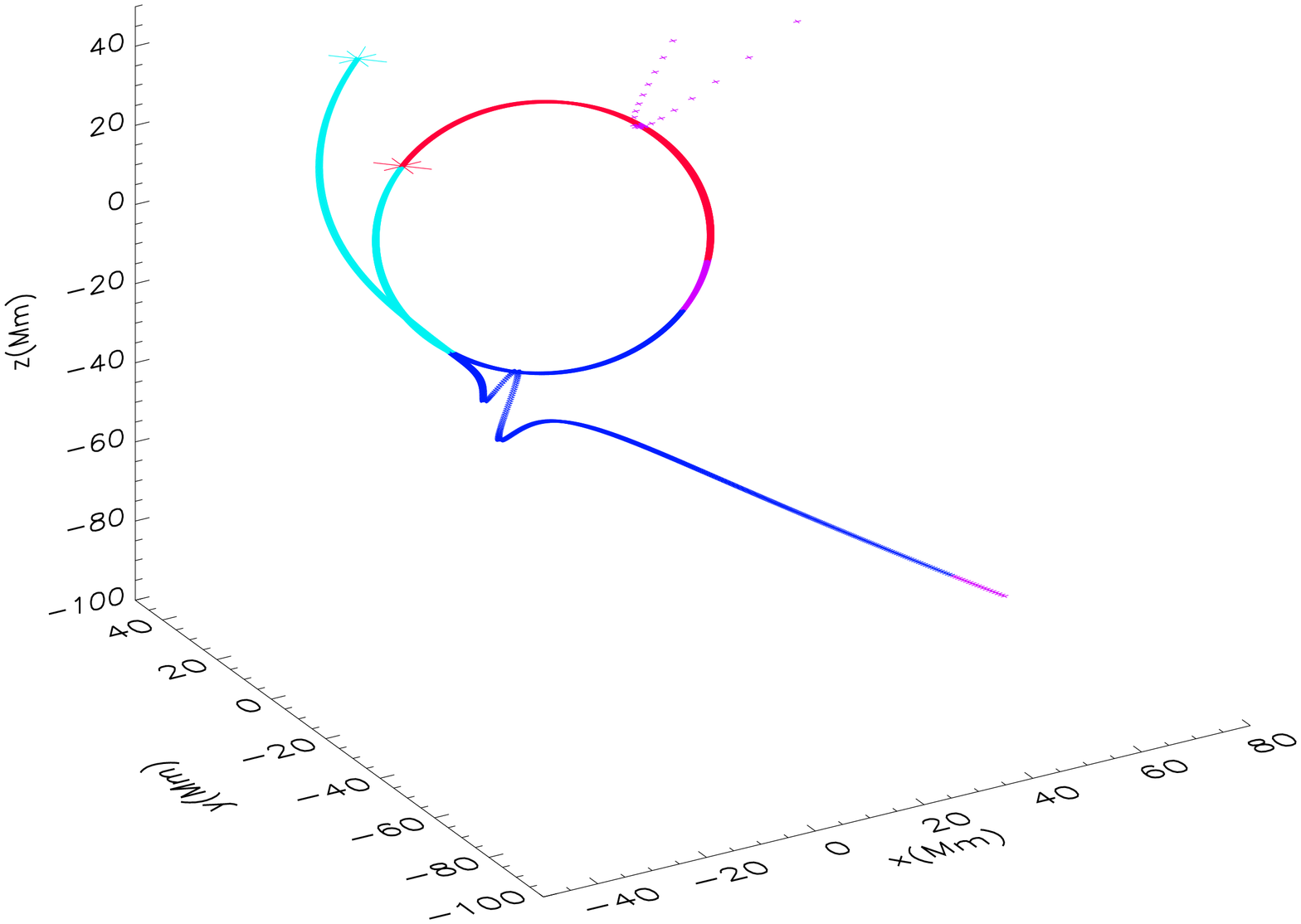}} 
\\
(i) 
{\label{fig:i}\includegraphics[width=0.65\textwidth,height=0.7\columnwidth] 
{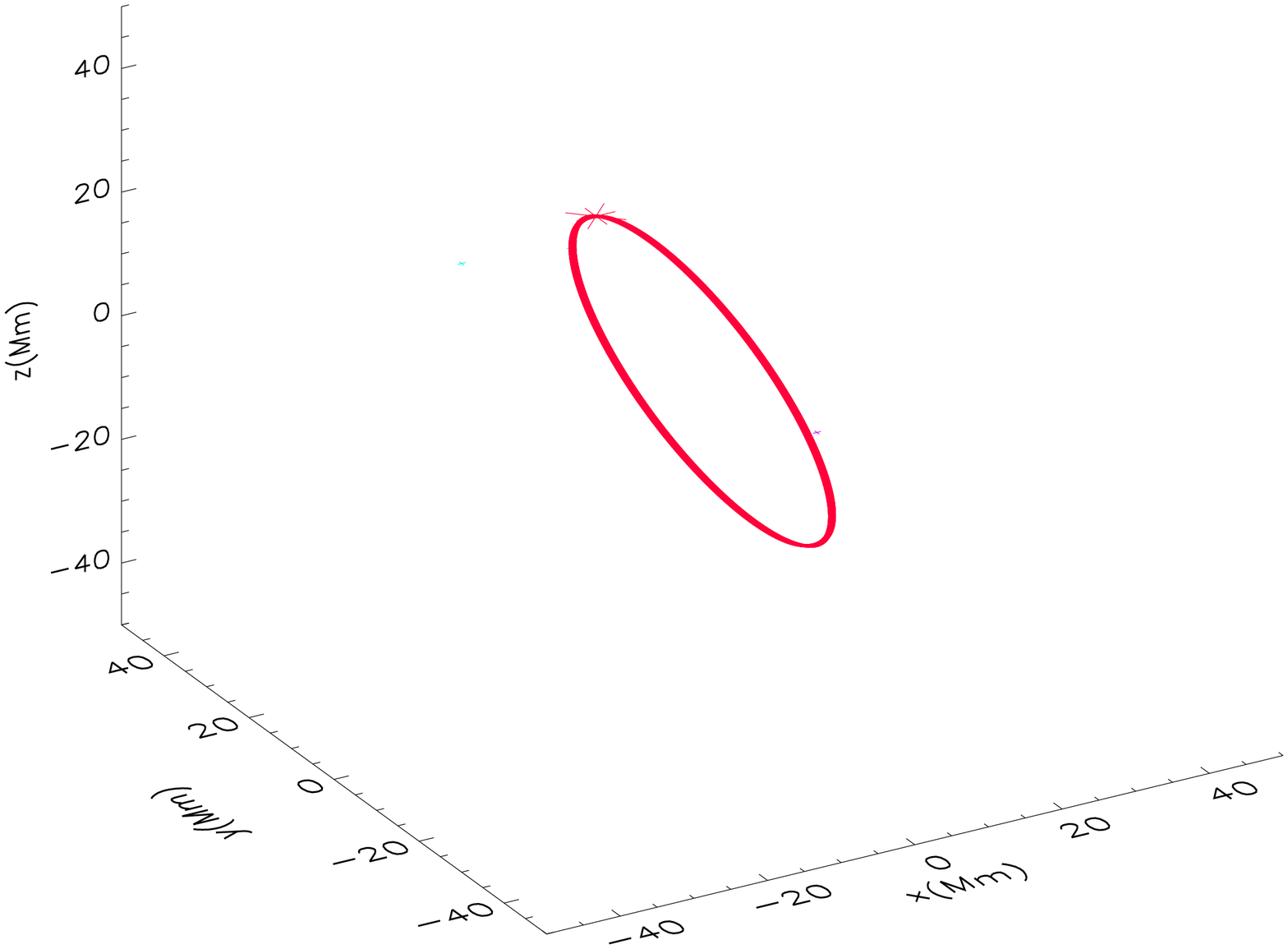}}  
\caption{Top: positioning a Galileo satellite with four GPS emitters.
Bottom: same for a GPS satellite and four Galileo emitters.} 
\label{fig4} 
\end{figure*}

In this paper, we are concerned with the location of users which move far away 
from Earth as, e.g., an user in a satellite. Of course, we use the emission times 
broadcast by four satellites, which might belong, e.g., to GPS or  
Galileo GNSSs. Two particular cases are considered. In the first (second) one,
the user travels in a Galileo (GPS) satellite and the emitters are four 
GPS (Galileo) satellites. Thus, the world lines of the user and the emitters
are known (see section~\ref{sec1}). As it follows from 
section~\ref{sec2}, there is no bifurcation for distances to $E$ smaller than about $10^{4} \ km$,
which means that GPS and Galileo satellites, which have altitudes of $20200$ 
and $23222 \ km$, may pass from a regions with bifurcation to other region
with single positioning or vice versa. The parts of the user circumference where bifurcation
occurs are now determined in the two cases. Results of the first (second) case are presented 
in panel (h) [(i)] of Fig.~(\ref{fig4}). 

7200 equally spaced points are considered on the user world line. In each
point, the emission coordinates are calculated (Newton-Rhapson) and, from them, the sign
of $\chi^{2} $ is found, this sign tell us if the point is single valued or it
has an associated false position (bifurcation).
Single valued points are red and are always located on the user circumference.

Four sets of 1800 points have been selected and --in the case of bifurcation-- these
sets have been ordered in the sense of growing time (dextrogyre) by using the following 
sequence of colours: black, fuchsia, dark blue and light blue.

Initial points may be: a single valued red point (represented by a star) or 
a bifurcation represented by two black stars.

Since GPS and GALILEO satellites have not the same period,
the final point has not always the same $\chi^{2} $ sign as the initial one. In the case 
of bifurcation one of the points is on the circumference and the other point 
is an external light blue star.

In the transition from red (single positioning) to any other colour (bifurcation),
one of the positions is on the circumference and the other one tends to infinity. The same 
occurs from any colour (bifurcation) to red (single positioning). Any other colour change is continuous.
It is due to the fact that we have decided to change the colour
to follow the satellite motion.

In panel (i) of Fig.~(\ref{fig4}), there are no bifurcation at all (red points).
This is an exceptional case. A more frequent situation with zones of bifurcation
is given in panel (h), where the asymptotic behaviour at the ends
of the bifurcation intervals is displayed.

\section{General discussion}

In our approach, satellites move in Schwarzschild space-time, so 
the effect of the Earth gravitational field on the clocks is taken into account,
e.g., it has been verified that GPS clocks run more rapid than clocks at rest on Earth by about 
38.4 microseconds per day. This prediction agrees with previous ones, which 
strongly suggests that our methods and codes work.

Since the Earth gravitational field produces a very small 
effect on photons while they travel from the satellites to the receiver
(the covered distance is not large and the gravitational field is weak),
photons have been moved in Minkowski space-time.  

We are currently moving photons in Schwarzschild, Kerr, 
and PPN space-times; however, only small corrections arise with respect 
to the approach assumed here. Previous work on this subject has been 
performed in various papers~\cite{cad10,del11,bun11}.

In this paper, the emission coordinates are calculated, from the inertial ones, 
by using accurate numerical codes based on the Newton-Raphson method.
However, the inertial coordinates are obtained, from the emission ones (positioning),
by means of the analytical transformation law derived in~\cite{coll10}.

From the emission coordinates and the satellite world lines,
one easily finds the number of possible receiver positions.
If this number is two, there is bifurcation. 
In this case, it has been proposed a method (based on angle measurements)
to select the true position~\cite{coll12}. Other methods (based on time measurements)
are possible (see~\cite{puc12}).

We have proved that small uncertainties in the satellite world lines produce
large positioning errors if $J \simeq 0$. A more detailed study 
of this type of errors is in progress.
 
The emission region has been studied for a certain 4-tuple of satellites.
The zones with bifurcation and those having small values of $|J| $ have 
been found, and appropriate methods have been used to their representation.
We have seen that satellites moving at altitudes greater than about 
$10^{4} \ km $ may cross these zones, which leads to problems due to 
bifurcation and large positioning errors.
In a GNSS there are various 4-tuples of satellites which may be used 
to find the position of a certain user. Among the possible 4-tuples 
without bifurcation, we should choose the 4-tuple leading to the 
greatest value of $|J| $ to minimize positioning errors.

Positioning on Earth surface is always single ($\chi^{2} \leq 0$) and the Jacobian 
does not vanish in this case. Hence, our study is particularly relevant
in the case of users moving
far away from Earth.

\vspace{0.5 cm} 
 
{\bf Acknowledgments}
This work has been supported by the Spanish
Ministries of {\em Ciencia e Innovaci\'on} and {\em Econom\'{\i}a y Competitividad},
MICINN-FEDER projects FIS2009-07705 and FIS2012-33582.

\end{document}